\documentclass[12pt]{article}
\usepackage{graphicx}
\begin{document}

\begin{center}
{\Large \bf Symmetries of the Poincar\'e sphere and decoherence matrices}

\vspace{5ex}

S. Ba{\c s}kal \footnote{electronic address:
baskal@newton.physics.metu.edu.tr} \\
Department of Physics, Middle East Technical University,
06531 Ankara, Turkey \\
\vspace{3ex}
Y. S. Kim \footnote{electronic address: yskim@physics.umd.edu}\\
Department of Physics, University of Maryland, College Park, Maryland 20742
\end{center}

\begin{abstract}
The Stokes parameters form a Minkowskian four-vector under various optical
transformations.  As a consequence, the resulting two-by-two density matrix
constitutes a representation of the Lorentz group.  The associated Poincar\'e
sphere is a geometric representation of the Lorentz group.
Since the Lorentz group preserves the determinant of the density matrix,
it cannot accommodate the decoherence process through the decaying
off-diagonal elements of the density matrix, which yields to an
incerese in the value of the determinant.
It is noted that the $O(3,2)$ deSitter group contains two Lorentz subgroups.
The change in the determinant in one Lorentz group can be compensated
by the other.  It is thus possible to describe the decoherence process
as a symmetry transformation in the $O(3,2)$ space.  It is shown also
that these two coupled Lorentz groups can serve as a concrete example of
Feynman's rest of the universe.
\end{abstract}

\vspace{30mm}


\newpage
\section{Introduction}\label{intro}

Traditionally the Poincar\'e sphere plays the central role in the
polarization optics~\cite{born80}.  The sphere is also applicable
to all two-beams systems with partially coherent phase
relations~\cite{cloude86,hkn00}.  The sphere has many interesting
symmetry properties.  Of course, this sphere has three-dimensional
rotational symmetries which are well known.  What other symmetries
does this sphere possesses?  This is the question we would like to
address in this paper.

Polarization optics can also be formulated in terms of the two-by-two
and four-by-four representations of the six-parameter Lorentz
group.  It was noted that the
two-component Jones vector and the four-component Stokes parameters
are like the relativistic spinors and the Minkowskian four-vectors,
respectively~\cite{hkn97,brown95}.  It is possible to identify
the attenuator, rotator,
and phase shifter with appropriate transformation matrices of the
Lorentz group.
This formulation is not restricted to polarization optics.  It
can be applied to all two-beam systems with coherent or partially
coherent phases.

If we use $(t, z, x, y)$ as the Minkowskian four-vector to which
four-by-four Lorentz-transformation matrices are applicable, it
is possible to write
\begin{equation}\label{mat22}
X = \pmatrix{t + z & x - iy  \cr x + iy & t - z} ,
\end{equation}
with appropriate two-by-two transformation matrices applicable
to both sides of this two-by-two representation of the four-vector.
These Lorentz transformations are unimodular transformations,
keeping the determinant of the above matrix constant.  We can write
this in the familiar form
\begin{equation}\label{det11}
t^2 -z^2 - x^2 - y^2 = constant .
\end{equation}

If we write the Stokes parameters in this two-by-two form, the
matrix becomes the density matrix.  This density matrix can also
be geometrically represented by the Poincar\'e sphere.
Therefore, the symmetry of the Poincar\'e sphere is necessarily
that of the Lorentz group~\cite{hkn00}.  In this Lorentzian regime,
the determinant of the density matrix is an invariant quantity.

Unlike the Jones vectors, the Stokes parameters, density matrix, and
the Poincar\'e sphere can deal with the lack of coherence between
the two beams.  The determinant of the density matrix vanishes when
the two beams are completely coherent, and it increases as the beams
lose coherence.  The Lorentzian symmetry of the Poincar\'e sphere can
describe the symmetry with a fixed value of the determinant, but it
cannot describe the process in which the determinant changes its value.
In other words, we cannot discuss the decoherence process within the
framework of the Lorentz group~\cite{hkn00}.

This decrease in coherence is an irreversible process, and we are
tempted to associate this problem with dissipation problems in
physics~\cite{gulen93}.
Of course, the mathematical method closest to group theoretical methods
is to introduce the concept of dissipative groups or
semi-groups~\cite{solo02}.   While this method is quite promising in
traditional dissipation problems, we choose take care of this
this decoherence problem with a mathematical method which is already
familiar to us.

Let us start with a pair of complex numbers $a$ and $b.$  From
these numbers, we can construct the density matrix of the form
\begin{equation}\label{den11}
\rho = \pmatrix{aa^* & ab^* e^{-\lambda t} \cr
        a^*b e^{-\lambda t}  & bb^*} .
\end{equation}
Indeed, the decay in the off-diagonal elements of this matrix plays
fundamental role in decoherence processes~\cite{zurek81,joos85}.

The determinant of this matrix is
\begin{equation}\label{det22}
  aa^*bb^*\left(1 - e^{-2\lambda t}\right) .
\end{equation}
This density matrix enjoys the symmetry properties like those
of the $X$ matrix given in Eq.(\ref{mat22}), since the
optical transformations applicable to the Stokes parameters are like
Lorentz transformations.  However, these determinant-preserving
transformations cannot change the $t$ variable.

When $t = 0$, the system is in a pure state, and the determinant is zero.
As $t$ increases, the value of the determinant in Eq.(\ref{det22})
increases from zero to $aa^*bb^*$, and consequently the system becomes
decoherent.

The question is whether there is a symmetry group which will
accommodate this transition process.  We know the Lorentz
group cannot, but this does not prevent us from looking for
a larger symmetry group.  The purpose of the present paper is
to show that the deSitter group $O(3,2)$ accommodates this decoherence
process.

This deSitter group is a Lorentz group applicable to a five-dimensional
space consisting of three space coordinates and two time coordinates.
While the three-dimensional rotation group is applicable to the
three space coordinates, the one-parameter two-dimensional rotation
group is applicable to the two time coordinates.

Although, this may sound like a mathematical exercise remote from the
physical reality, we would like emphasize that the $O(3,2)$ deSitter group
is already a standard theoretical tool in optical sciences,
specifically as a mathematical basis for two-mode squeezed
states~\cite{vourdas88,hkn90}, as well as in the theory of elementary
particles together with the $O(4,1)$ group.  As Paul A. M. Dirac noted
in 1963, the $O(3,2)$ group is the fundamental symmetry group for two
coupled harmonic oscillators~\cite{dir63}.  This two-oscillator system
often serves as a  mathematical basis for soluble models such as the Lie
model in quantum field theory~\cite{sss61} and the Bogoliubov
transformations in superconductivity~\cite{fewa71}.

In this paper we are interested in the fact that the $O(4,1)$ group
contains two $O(3,1)$ Lorentz groups, where the two time variables are
linearly combined through the one parameter rotation group.
We will consider them as two coupled Lorentzian spaces.
The loss of coherence in one Lorentzian space will result in the gain in
the other space.  We shall show that our symmetry model will constitute
a concrete example of Feynman's rest of the universe.  The first Lorentzian
space is the world in which we make physical observations, and the second
space belongs to the rest of the universe~\cite{fey72,hkn99ajp}.

In Sec.~\ref{minkow}, we review the symmetries of the Stokes parameters
and the density matrix.  In Sec.~\ref{poins}, we study the symmetries of
the Poincar\'e sphere within the Lorentzian framework and discuss in
detail what is possible and what is not possible.  In Sec.~\ref{o32},
it is shown that the $O(3,2)$ symmetry can provide a framework for
the decoherence process.  In Sec~\ref{frest}, we interpret the result
of our paper in terms of Feynman's rest of the universe.

\section{Stokes Parameters as Minkowskian Four-vectors}\label{minkow}
Let us start with a plane wave propagating along the $z$ direction.
Then, it has polarizations along the $x$ and $y$ directions.  We can
then write the Jones vector as
\begin{equation}\label{jones}
\pmatrix{\psi_{1} \cr \psi_{2}} =
\pmatrix{A \exp{\left\{i(kz - \omega t)\right\}} \cr
B \exp{\left\{i(kz - \omega t)\right\}} } .
\end{equation}
Even though the Jones vector was developed originally for
polarized light waves, the formalism can be extended to all
two-beam systems such as interferometers~\cite{hkn00}.

If the two beams are mixed, we use the rotation matrix
\begin{equation}\label{rot22}
R(\theta) = \pmatrix{\cos(\theta/2) & -\sin(\theta/2) \cr
\sin(\theta/2) & \cos(\theta/2) } ,
\end{equation}
applicable to column vector of Eq.(\ref{jones}).

These two beams can go through two different optical path lengths,
resulting in a phase difference.  If the phase difference is
$\phi$, the phase shift matrix is
\begin{equation}\label{shif22}
P(\phi) = \pmatrix{e^{-i\phi/2} & 0 \cr 0 & e^{i\phi/2}} .
\end{equation}

When reflected from mirrors, or while going through beam splitters,
there are intensity losses for both beams.  The rate of loss is not
the same for the beams.  This results in the attenuation matrix
of the form
\begin{equation}\label{atten}
\pmatrix{e^{-\eta_{1}} & 0 \cr 0 & e^{-\eta_{2}}} =
e^{-(\eta_{1} + \eta_{2})/2} \pmatrix{e^{\eta/2} & 0 \cr 0 &
e^{-\eta/2}}
\end{equation}
with $\eta = \eta_{2} - \eta_{1}$ .
This attenuator matrix tells us that the electric fields are
attenuated at two different rates.  The exponential factor
$e^{-(\eta_{1} + \eta_{2})/2}$ reduces both components at the same
rate and does not affect the degree of polarization.  The effect of
polarization is solely determined by the squeeze matrix
\begin{equation}\label{sq22}
S(\eta) = \pmatrix{e^{\eta/2} & 0 \cr 0 & e^{-\eta/2}} .
\end{equation}

It was shown in Refs.~\cite{hkn97,hkn00} that repeated applications of the
rotation matrices of the form of Eq.(\ref{rot22}), shift matrices of
the form of Eq.(\ref{shif22}) and squeeze matrices of the form of
Eq.(\ref{sq22}) lead to a two-by-two representation of the six-parameter
Lorentz group.  The transformation matrix in general takes the form
\begin{equation}\label{lt22}
G = \pmatrix{\alpha & \beta \cr \gamma & \delta} ,
\end{equation}
applicable to the column vector of Eq.(\ref{jones}), where all four
elements are complex numbers with the condition that the determinant
of the matrix be one.  This matrix contains six free parameters.
The above $G$ matrix constitutes the two-by-two representation of the
six-parameter Lorentz group, commonly called $SL(2,c)$.

Indeed, the two-component Jones vector provides the representation
space for the two-by-two representation of the Lorentz group.  However,
the Jones vectors cannot describe whether the two-beams are coherent.
This is the reason why we have to resort to the coherency matrix
\begin{equation}\label{cocy11}
C = \pmatrix{S_{11} & S_{12} \cr S_{21} & S_{22}},
\end{equation}
with
\begin{eqnarray}\label{sii}
&{}& S_{11} = <\psi_{1}^{*}\psi_{1}>  , \qquad
S_{22} = <\psi_{2}^{*}\psi_{2}> , \nonumber \\[2ex]
&{}& S_{12} = <\psi_{1}^{*}\psi_{2}> ,  \qquad
S_{21} = <\psi_{2}^{*}\psi_{1}> .
\end{eqnarray}
This coherency matrix also serves as the density matrix~\cite{fey72}.

Under the influence of the $G$ transformation given in Eq.(\ref{lt22}),
this density matrix is transformed as
\begin{eqnarray}\label{trans22}
&{}& C' = G\,C\,G^{\dagger} =
\pmatrix{S'_{11} & S'_{12} \cr S'_{21} & S'_{22}} \nonumber \\[2ex]
&{}&\hspace{5ex} = \pmatrix{\alpha & \beta \cr \gamma & \delta}
\pmatrix{S_{11} & S_{12} \cr S_{21} & S_{22}}
\pmatrix{\alpha^{*} & \gamma^{*} \cr \beta^{*} & \delta^{*}} .
\end{eqnarray}
This leads to the four-by-four transformation
\begin{equation}\label{trans44}
\pmatrix{S_{11}' \cr S_{22}' \cr S_{12}' \cr S_{21}'} =
\pmatrix{
\alpha^{*}\alpha & \gamma^{*}\beta &
\gamma^{*}\alpha & \alpha^{*}\beta \cr
\beta^{*}\gamma  & \delta^{*}\delta &
\delta^{*}\gamma & \beta^{*}\delta \cr
\beta^{*}\alpha  & \delta^{*}\alpha &
\beta^{*}\beta   & \delta^{*}\beta  \cr
\alpha^{*}\gamma & \gamma^{*}\gamma &
\alpha^{*}\delta & \gamma^{*}\delta }
\pmatrix{S_{11} \cr S_{22} \cr S_{12} \cr S_{21}}  .
\end{equation}

It is sometimes more convenient to use the following combinations of
parameters.
\begin{eqnarray}\label{stokes}
&{}& S_{0} = \frac{S_{11} + S_{22}}{\sqrt{2}},  \qquad
    S_{1} = \frac{S_{11} - S_{22}}{\sqrt{2}},    \nonumber \\[2ex]
&{}& S_{2} = \frac{S_{12} + S_{21}}{\sqrt{2}}, \qquad
S_{3} = \frac{S_{12} - S_{21}}{\sqrt{2} i} .
\end{eqnarray}
These four parameters are called the Stokes parameters in the
literature~\cite{shur62}, usually in connection with polarized
light waves.  However, as was mentioned before, the Stokes parameters
are useful to all two-beam systems.  We can write the above expression
as
\begin{equation}
\pmatrix{S_{0} \cr S_{1} \cr S_{2} \cr S_{3}} =\frac{1}{\sqrt{2}}
\pmatrix {(S_{11} + S_{22}) \cr (S_{11} - S_{22}) \cr
(S_{12} + S_{21}) \cr i(S_{21} - S_{12}) } .
\end{equation}
Then the four-by-four matrix which transforms
$\left(S_{11}, S_{22}, S_{12}, S_{21} \right)$ to
$\left(S_{0}, S_{1}, S_{2}, S_{3} \right)$ is
\begin{equation}
\pmatrix{S_{0} \cr S_{1} \cr S_{2} \cr  S_{3} } = \frac{1}{\sqrt{2}}
 \pmatrix{1 & 1 & 0 & 0 \cr 1 & -1 & 0 & 0 \cr
 0 & 0 & 1 & 1 \cr  0 & 0 & -i & i}
\pmatrix{S_{11} \cr S_{22} \cr S_{12} \cr  S_{21} } .
\end{equation}
This matrix enables us to construct the transformation matrix
applicable to the Stokes parameters, widely known as the Mueller
matrix.
The transformation matrix applicable to the Stokes parameters of
Eq.(\ref{stokes}) can be derived from Eq.(\ref{trans44}), and its
form has been discussed in detail in Refs.~\cite{hkn00,hkn97}.
The above Stokes parameters form a Minkowskian
four-vector
like $(t, z, x, y)$, and the transformation matrix applicable to
the Stokes parameters represents a Lorentz transformation.

The four-by-four representation  is like the Lorentz transformation
matrix applicable to the space-time Minkowskian vector
$(t, z, x, y)$~\cite{hkn00}.
This allows us to study space-time symmetries in terms of the Stokes
parameters which are applicable to interferometers.
Let us first see how the rotation matrix of Eq.(\ref{rot22}) is
translated into the four-by-four formalism.  In this case,
\begin{equation}
\alpha = \delta = \cos(\theta/2), \qquad
\gamma = -\beta = \sin(\theta/2) .
\end{equation}
Thus, the corresponding four-by-four matrix takes the form
\begin{equation}\label{rot44}
R(\theta) = \pmatrix{1 & 0 & 0 & 0 \cr
0 & \cos\theta & -\sin\theta & 0  \cr
0 & \sin\theta & \cos\theta & 0 \cr
0 & 0 & 0 & 1} .
\end{equation}

Let us next see how the phase-shift matrix of Eq.(\ref{shif22}) is
translated into this four-dimensional space.  For this two-by-two
matrix,
\begin{equation}
\alpha = e^{-i\phi/2} , \qquad \beta = \gamma = 0 , \qquad
\delta = e^{i\phi/2} .
\end{equation}
For these values, the four-by-four transformation matrix
takes the form
\begin{equation}\label{shif44}
P(\phi) = \pmatrix{1 & 0 & 0 & 0 \cr 0 & 1 & 0 & 0  \cr
0 & 0 & \cos\phi & -\sin\phi \cr 0 & 0 & \sin\phi & \cos\phi} .
\end{equation}
For the squeeze matrix of Eq.(\ref{sq22}),
\begin{equation}
\alpha = e^{\eta/2}, \qquad \beta = \gamma = 0 , \qquad
\delta = e^{-\eta/2} .
\end{equation}
As a consequence, its four-by-four equivalent is
\begin{equation}\label{sq44}
S(\eta) = \pmatrix{\cosh\eta & \sinh\eta & 0 & 0 \cr
\sinh\eta & \cosh\eta & 0 & 0 \cr
0 & 0 & 1 & 0 \cr 0 & 0 & 0 & 1} .
\end{equation}
If the above matrices are applied to the four-dimensional
Minkowskian space of $(t, z, x, y)$, the above squeeze matrix will
perform a Lorentz boost along the $z$ or $S_{1}$ axis with $S_{0}$ as
the time variable.  The rotation matrix of Eq.(\ref{rot44}) will
perform a rotation around the $y$ or $S_{3}$ axis, while the phase
shifter of Eq.(\ref{shif44}) performs a rotation around the $z$ or
the $S_{1}$ axis.  Matrix multiplications with $R(\theta)$ and
$P(\phi)$ lead to the three-parameter group of rotation matrices
applicable to the three-dimensional space of $(S_{1}, S_{2}, S_{3})$.

The phase shifter  $P(\phi)$ of Eq.(\ref{shif44}) commutes with the
squeeze matrix of Eq.(\ref{sq44}), but the rotation matrix $R(\theta)$
does not.  This aspect of matrix algebra leads to many interesting
mathematical identities which can be tested in laboratories.  One of
the interesting cases is that we can produce a rotation by performing
three squeezes.  This aspect is widely known as the Wigner rotation as
discussed in the literature.

In this paper, we are interested in studying the time-dependent
density matrix of the form
\begin{equation}\label{cocy22}
C(t) = \pmatrix{S_{11} & S_{12}e^{-\lambda t} \cr
S_{21} e^{-\lambda t} & S_{22}}.
\end{equation}
This matrix can be translated into the Minkowskian four-vector
\begin{equation}
\pmatrix{S_{0} \cr S_{1} \cr S_{2} e^{-\lambda t} \cr S_{3}e^{-\lambda t} } .
\end{equation}
As $t$ increases, the third and fourth component of this Minkowskian
four-vector becomes smaller.

Lorentz transformations preserve the $(length)^2 $ of the four-vector
which in the Minkowskian metric takes the form
\begin{equation}
S_{0}^2 - S_{1}^2 - (S_{2}^2  +  S_{3}^2) e^{-2\lambda t} .
\end{equation}
This is also the determinant of the density matrix $D(t)$.  If this
quantity increases as the time $t$ increases, we cannot handle the
problem within  the framework of the Lorentz group~\cite{hkn00}.

One option is to assert that this is not a reversible problem and
invent a mathematical tool other than group theory~\cite{solo02}.
Another approach is to look for a larger group which contains the
Lorentz group as a subgroup.  This is precisely what we intend to do
in this paper.  In Sec.~\ref{o32}, we shall introduce the $O(3,2)$
deSitter group which contains two Lorentz groups.  Before getting
into the world of the $O(3,2)$ symmetry, let us study the geometry
of the Poincar\'e sphere in the following section.

\section{Lorentz Symmetries of the Poincar\'e Sphere}\label{poins}

The Poincar\'e sphere has a long history, and its spherical symmetry is
well known~\cite{born80}.  The Lorentz group has the three-dimensional
rotation group as its subgroup.  Thus, the Lorentz symmetry of the
Poincar\'e sphere includes the traditional rotational symmetry.  Let
us study in this section the symmetries associated with Lorentz boosts.

If we use the expressions of $\psi_1$ and $\psi_2$ given in Eq.(\ref{jones}),
the density matrix $C$ of Eq.(\ref{cocy11}) becomes
\begin{equation}\label{den22}
D(t) = \pmatrix{A^2 & AB e^{(-\lambda t - i\phi)} \cr
 AB e^{(- \lambda t + i\phi)} & B^2}.
\end{equation}
Here $\phi$ is the phase difference between
$\psi^*_{1}\psi_{2}$ and $\psi_{1}\psi^*_{2}$.  The $\lambda t$ factor
in the exponent describes the loss of coherence.  We assume that the
off-diagonal terms decrease exponentially in the time variable.  The
determinant of this density matrix is
\begin{equation}\label{det55}
(AB)^{2} \left(1 - e^{-2\lambda t}  \right) .
\end{equation}
This determinant is zero when $t = 0$, but increases to $(AB)^2$ as
$t$ becomes larger.

The corresponding four-vector is
\begin{equation}\label{4vec33}
\frac{1}{\sqrt{2}} \pmatrix{A^2 + B^2  \cr A^2 - B^2 \cr
2AB(\cos\phi)e^{-\lambda t} \cr 2AB(\sin\phi)e^{-\lambda t} } .
\end{equation}
For a fixed value of $t$, the geometry of the Poincar\'e sphere is the
geometry defined by the three parameters $A, B$ and $\phi$.
This sphere consists of two spheres:
One is the outer sphere whose radius is the time-like component
of the above four-vector
\begin{equation}
s =\frac{(A^2 + B^2)}{2} ,
\end{equation}
and the other is the inner sphere whose radius is the
magnitude of the three-vector contained in the four-vector of
Eq.(\ref{4vec33})
\begin{equation}
r =\frac{1}{2} \sqrt{\left(A^2 - B^2\right)^2 + 4 (AB)^2 e^{-2\lambda t} } .
\end{equation}
Then the quantity
\begin{equation}
s^2 - r^2
\end{equation}
is Lorentz-invariant, and is equal to the value of the determinant given
in Eq.(\ref{det55}).  The inner radius is equal to the outer radius when
$t = 0$, and becomes $\left(A^2 - B^2\right)/2$ as $t$ becomes very large.

We can now introduce a spherical coordinate system with
\begin{eqnarray}
&{} r_{z} = (A^2 - B^2)/2  = r (\cos\theta) , \nonumber \\[2ex]
&{} r_{x} =  AB(\cos\phi)e^{-\lambda t} =
                        r (\sin\theta) \cos\phi, \nonumber \\[2ex]
&{} r_{y} =  AB(\sin\phi)e^{-\lambda t} = r (\sin\theta) \sin\phi .
\end{eqnarray}
Then the Lorentz symmetry allows rotations in this three-dimensional
system.  Now, with the appropriate rotation it is possible to bring
four-vector of Eq.(\ref{4vec33}) to
\begin{equation}\label{4vec44}
\pmatrix{s \cr r \cr 0 \cr 0}.
\end{equation}
The rotations do not change the radii of the outer and inner spheres,
and $r$ and $s$ remain invariant under the rotations.

However, the Lorentz symmetry allows the Lorentz boosts of
the four-vector of Eq.(\ref{4vec44}) along the $-z$ direction.  If we apply
the inverse of the boost matrix of Eq.(\ref{sq44}), then the four-vector
becomes
\begin{equation}\label{4vec55}
\pmatrix{s(\cosh\eta) - r (\sinh\eta) \cr r (\cosh\eta)
- s (\sinh\eta) \cr 0 \cr 0 }.
\end{equation}
This transformation changes the outer and inner radii, but
keeps $(s^2 - r^2)$ invariant, as we can see from
\begin{equation}
[s(\cosh\eta) - r (\sinh\eta)]^2 -
[r (\cosh\eta)- s (\sinh\eta)]^2 = s^2 - r^2 .
\end{equation}
It is now possible to choose the
value of $\eta$ such that
\begin{equation}
r (\cosh\eta)- s (\sinh\eta) = 0 ,
\end{equation}
which leads to $\tanh\eta = r/s$.  If this condition is met, the
four-vector of Eq.(\ref{4vec55}) becomes
\begin{equation}\label{4vec66}
\pmatrix{\sqrt{s^2 - r^2} \cr 0 \cr 0 \cr 0} =
\pmatrix{AB\sqrt{1 - e^{-2\lambda t}} \cr 0 \cr 0 \cr 0} .
\end{equation}
Indeed, the Lorentz symmetry allows us to bring the Poincar\'e sphere
to a one-number system.  We are now tempted to change the value of
$\left(r^2 - s^2\right)$ in the above expression by changing the
time variable $t.$  This is precisely what is not allowed within
the framework of the Lorentz group.  We shall see whether this
can be achieved when symmetry group is enlarged.

\section{O(3,2) Symmetry of the Poincar\'e Sphere}\label{o32}
In order to deal with this problem, we introduce the $O(3,2)$
deSitter space with $(t, z, x, y, u)$ where $t$ and $u$ are two time-like
variables while allowing two-dimensional rotations in the $t$ and $u$.
As we emphasized in Sec.~\ref{intro}, this group has already been
exploited in optical sciences.  For instance, it is the fundamental
language for two-mode squeezed states~\cite{vourdas88,hkn90}.

In this deSitter space, we are allowed to have the rotation
\begin{equation}\label{tu}
\pmatrix{\cos\chi & 0 & 0 & 0 & \sin\chi  \cr
0 & 1 & 0 & 0 & 0 \cr 0 & 0 & 1 & 0 & 0 \cr
0 & 0 & 0 & 1 & 0 \cr -\sin\chi & 0 & 0 & 1 & \cos\chi}
\pmatrix{0 \cr 0 \cr 0 \cr 0 \cr m} =
\pmatrix{m\,(\sin\chi) \cr 0 \cr 0 \cr 0 \cr m\,(\cos\chi)} .
\end{equation}
Now, the invariant quantity is
\begin{equation}
t^2 + u^2 - z^2 - x^2 - y^2.
\end{equation}
As we can see from Eq.(\ref{tu}), if $z = x = y = 0$, this quantity is
$(t^2+ u^2) = m^2$, and remains as an invariant in this space.
The deSitter space contains two Minkowskian subspaces, namely the spaces of
$(t, z, x, y)$ with the invariant of $\left(t^2 - z^2 - x^2 - y^2\right)$,
and of $(u, z, x, y)$ with the invariant of
$\left(u^2 - z^2 - x^2 - y^2\right)$.

Let us consider the five-vector $(0, 0, 0, 0, m)$ in this space.
The above five-by-five matrix changes this five-vector to
\begin{equation}
\left(m\,(\sin\chi), 0, 0, 0, m\,(\cos\chi)\right) .
\end{equation}
Thus, in the Minkowskian world of $(t, z, x, y)$, the invariant quantity is
$m^2\sin^2\chi$, and $m^2\cos^2\chi$ in the Minkowskian space of
$(u, z, x, y)$, where now the four vectors in these spaces are
\begin{equation}\label{4vec77}
\pmatrix{m \,(\sin\chi) \cr 0 \cr 0 \cr 0 } , \qquad
\pmatrix{m \,(\cos\chi) \cr 0 \cr 0 \cr 0 }
\end{equation}
respectively.

Let us compare the first four-vector of Eq.(\ref{4vec77}) with the four-vector of
Eq.(\ref{4vec66}).  If we identify the parameter $m (\sin\chi) $ in
Eq.(\ref{4vec77}) with $\sqrt{s^2 - r^2}$ of Eq.(\ref{4vec66}), we have
\begin{equation}
s^2 - r^2 = m^2 \sin^2\chi .
\end{equation}
This further allows us to identify $m$ as $AB$ in Eq.(\ref{4vec66}), and
\begin{equation}
(AB)^2(\sin\chi)^2 = (AB)^2 \left(1 - e^{-2\lambda t} \right) ,
\end{equation}
which leads to
\begin{equation}
\cos\chi = e^{-\lambda t} .
\end{equation}

We concluded in Sec.~\ref{poins} that the $t$ parameter cannot be changed
in the Lorentzian regime.  However, we have shown that this decoherence
parameter can be identified with the angle variable $\chi$ in the
deSitter space.

After changing the $t$ variable, we can make inverse transformations to
return to the four-vector of the form given in Eq.(\ref{4vec33}).  Indeed,
it is gratifying to note that we now have the freedom of changing this
time variable with a symmetry operation.  In terms of this symmetry
parameter, we can write the density matrix as
\begin{equation}\label{den66}
\rho(\chi) = \pmatrix{A^2 & AB\,e^{-i\phi}(\cos\chi)
\cr  AB\,e^{i\phi}(\cos\chi) &  B^2} .
\end{equation}
If $\chi = 0$ and $t= 0$, the system is in a pure state.  As $t$ becomes
large, the angle $\chi$ approaches $90^o.$  Therefore the
deSitter parameter $\chi$ neatly takes care of the loss of coherence
in the two-beam system.

\section{Feynman's Rest of the Universe}\label{frest}

In this paper, we insinuated two separate Minkowskian spaces
by introducing the deSitter space.  The first Minkowskian space
was defined by the coordinate variables $(t, z, x, y)$, and the
second one by $(u, z, x, y).$   When we discussed the Lorentzian
symmetry of the Poincar\'e sphere we worked with the first
Minkowskian space.   How about the second space?

Our analysis would be exactly the same, except that $\sin\chi$ is
replaced by $\cos\chi$ as can be seen from Eq.(\ref{tu}).  The density
matrix in this second space can then be written as
\begin{equation}\label{den77}
\sigma(\chi) = \pmatrix{A^2 & AB\,e^{-i\phi}(\sin\chi)
\cr AB\,e^{i\phi}(\sin\chi) & B^2} .
\end{equation}
This density matrix gains coherence as the density matrix of
Eq.(\ref{den66}) loses coherence.  The determinants of these two density
matrices are $(AB)^2\sin^2\chi $ and $(AB)^2\cos^2\chi$ respectively.
The sum of these two determinants is $(AB)^2$ and is independent of
angle variable $\chi.$  Indeed, these two density matrices or the two
Lorentzian subspaces are ``coupled'' in a Pythagorean manner.   What
is the meaning of this?

In his book on statistical mechanics~\cite{fey72}, Feynman makes
the following statement about the density matrix. 
{\it When we solve a quantum-mechanical problem, what we
really do is divide the universe into two parts - the system in which we
are interested and the rest of the universe.  We then usually act as if
the system in which we are interested comprised the entire universe.
To motivate the use of density matrices, let us see what happens when we
include the part of the universe outside the system}.

In order to understand what Feynman said, Han {\it et al.} used two
coupled oscillators to illustrate Feynman's rest of the
universe~\cite{hkn99ajp}.  One of the oscillators is in the world 
where we make measurements, and the other serves as the rest of 
the universe.  The two coupled oscillators form the entire universe.

By working with two separate Lorentz subgroups of the deSitter group,
we divided the universe into the world where we measure the degree
of decoherence and the hidden world which is still controlling the events 
in its counterpart.  The $O(3,2)$ deSitter world constitutes the entire 
universe.

It is gratifying to note that the present paper provides another
illustrative example of Feynman's rest of the universe.

\section*{Concluding Remarks}
It has been widely believed that the decoherence problem could not be
treated as a symmetry problem.  In this paper, we have presented a
different view, using an extra time-like dimension in the Lorentz group.
The deSitter group we used has been one of the standard tools in
relativistic quantum mechanics~\cite{salec58} and elementary particle
physics including one of the most recent models in string
theory~\cite{becker04}. Also, this group is not new in optical sciences.  
In 1963, Paul A. M. Dirac observed that the deSitter group $O(3,2)$ 
serves as a symmetry group for coupled harmonic oscillators~\cite{dir63}.  
This group is the fundamental scientific language for two-mode squeezed 
states of light~\cite{vourdas88,hkn90}.  We are thus not carrying the burden 
of introducing a new mathematical device in this paper.

Of course, a more challenging problem is to compute the decay parameter
$\lambda$ from dynamical considerations, but this is beyond the scope
of the present paper dealing solely with symmetry problems.  However,
this symmetry property may be helpful in formulating dynamical problems
in the future.

As we noted in Sec.~\ref{frest}, the $O(3,2)$ group can serve as an
illustrative example of Feynman's rest of the universe.  One Lorentz
subgroup represents the system under examination, while the other 
appears as the rest of the universe.  As Feynman noted, it is more 
satisfying to understand the entire system including the rest of the 
universe.

\section{Acknowledgments}

We would like to thank A. I. Solomon for explaining to us what
the semi-group is and how it can be applied to dissipation
problem in quantum mechanics.  We thank also N. L. Harshman
for interesting discussions on the semi group.

\end{document}